# Unveiling the potential of NdPO$_4$ magnetocaloric phases in cryogenic refrigeration

M. Balli[1]*, L. Attou[1], S-E. Bouzarmine[1, 3], S. Oubad[1], K. El Maalam[4], P. Fournier[2] and S. Mangin[3]

[1] International University of Rabat, Parc Technopolis, Rocade de Rabat-Salé, 11100, Morocco.
*mohamed.balli@uir.ac.ma
[2] Institut Quantique, Université de Sherbrooke, J1K 2R1, QC, Canada.
[3] Université de Lorraine, CNRS, IJL, F-54000 Nancy, France.
[4] MAScIR, Mohammed VI Polytechnic University, Lot 660, Hay Moulay Rachid Ben Guerir, Morocco

The RPO$_4$ orthophosphates (R = rare earth element) have recently attracted a wide interest due to the strong coupling between their electronic, orbital and structural ordering parameters resulting in a variety of functional properties. Herein, we demonstrate that NdPO$_4$ phases synthesized via a facile precipitation growth process unveil promise in low-temperature magnetic cooling. The analysis of their structural properties reveals nanorod forms with diameters of 10 to 20 nm and lengths ranging from 200 to 500 nm while the heat treatment transforms their hexagonal rhabdophane-type structure to a monoclinic anhydrous monazite-type symmetry. Magnetization measurements and DFT calculations show strong antiferromagnetic couplings and the absence of any magnetic ordering in the 2-300 K range. On the other hand, the monoclinic phase of NdPO$_4$ exhibits a large magnetocaloric effect of about 19 J/kg K under 5 T near 3 K, outperforming some reference materials containing more expensive rare-earth elements with high magnetic moments.

## I. INTRODUCTION

IN RECENT years, numerous materials have been reported to exhibit a giant magnetocaloric effect (GMCE) at cryogenic temperature range including both oxides and intermetallics [1-5]. However, the gap to be bridged in going from laboratory samples to competitive applications that meets the market needs is demanding. In fact, the magnetocaloric material must answer a series of requirements before its direct implementation such as sufficiently large magnetocaloric effect (MCE), low hysteresis, high resistance against oxidation and corrosion, mechanical stability and affordable constituent elements. In this context, the RPO$_4$ phosphates could be of great interest in low-temperature magnetic cooling because of their high chemical stability, strong mechanical stability and large magnetic moment at temperatures below 50 K [6, 7]. These materials have attracted a wide interest over the last decades because of their fascinating physical properties [6-9] as well as their high potential in several applications such as phosphors, lasers, nuclear waste and magnetic cooling [6-9]. Regarding the latter, a giant magnetocaloric effect has been recently reported in GdPO$_4$ single crystals [8]. This family of oxides [6, 7] usually crystallizes in two crystallographic structures which are the Zircon type (tetragonal) having the space group of D$_{4h}$ for R = Tb-Lu and the Monazite (monoclinic) with the space group of C$_{2h}$ for R = La-Gd. Herein, we investigate the potential of NdPO$_4$ in low-temperature magnetic cooling applications while shedding the light on the correlation between its structural, magnetic and magnetocaloric features.

## II. METHODS

Neodymium orthophosphates (NdPO$_4$) were synthesized through a facile precipitation route [9], using high-purity Nd(NO$_3$)$_3$.6H$_2$O (Sigma Aldrich), which served as the rare-earth precursor and was accurately weighed and dissolved in phosphoric acid (H$_3$PO$_4$) under continuous stirring. To ensure complete dissolution and minimize potential challenges posed by the viscosity of the acid, 3 mL of deionized water was incorporated into the mixture to enhance the fluidity of the solution. The latter was left to react overnight under continuous stirring to ensure complete complexation of the neodymium ions with the phosphate species. Subsequently, the reaction mixture was diluted with deionized water to a final volume of 100 mL and transferred to a round-bottom flask. The solution was refluxed at 130°C for two hours. The resulting precipitate was then carefully collected by centrifugal filtration, thoroughly rinsed with deionized water to eliminate residual impurities, and dried at 60 °C. This led to an intermediate phase (NdPO$_4$.H$_2$O in its hydrated form), which was then thermally annealed in air at 900°C for approximately 2 h to get the stable monoclinic monazite phase NdPO$_4$. The structural and microstructural properties of synthesized NdPO$_4$ phases were investigated by using high-resolution transmission electron microscopy and XRD measurements. The latter were carried out with the help of a Bruker D8 ADVANCE diffractometer with Cu-K$\alpha$ radiation over the 2$\theta$-angular range of 10° < 2$\theta$ < 90° with a step size of 0.01°. The hydrated and anhydrous phases of NdPO$_4$ were also analyzed by using complementary techniques such as thermogravimetric analysis (TGA), differential scanning calorimetry (DSC) and, attenuated Total Reflection-Fourier Transform Infrared (ATR-FTIR) spectroscopy. The magnetic properties were measured over the temperature range from 2 to 300 K under magnetic fields ranging between 0 and 7 T by using a superconducting quantum interference device (SQUID) magnetometer from Quantum Design (MPMS XL). Our experimental analyses were supported by calculations based on the Density Functional Theory (DFT) [10].

## III. RESULTS AND DISCUSSION

The results obtained demonstrate that the NdPO$_4$ is sensitive to the heat treatment conditions. In fact, the hydrated phase (NdPO$_4$·H$_2$O) obtained when subjecting the samples to a heat treatment under 400 °C forms in the hexagonal rhabdophane phase structure with the P3$_1$2 symmetry. In contrast, the anhydrous NdPO$_4$ phase which is treated under 900 °C crystallizes in a Monazite-type monoclinic structure within the space group P2$_1$/n (Fig 1-a). A detailed analysis of the microstructure reveals that both anhydrous and hydrated NdPO$_4$ phases form nanorods clustering into larger bundles with lengths usually comprised between 200 and 500 nm and diameters going from 10 to 20 nm (Fig 1-b).

The temperature dependence of magnetization under low magnetic fields for both phases reveals a similar magnetic behavior where magnetization markedly increases at low temperatures revealing the absence of any magnetic phase transition down to 2 K. This is consistent with the magnetic ordering temperature of about 0.32 K determined from DFT calculations based on the nearest-neighbor exchange (Inset Fig1-c) energy [10]. On the other hand, by linearly fitting the inverse magnetic susceptibility with the Curie-Weiss law at high temperatures, effective magnetic moments ($\mu_{eff}$) and Curie-Weiss temperatures (T$_\theta$) were found to be 3.86 $\mu_B$ and -89 K for the hydrated NdPO$_4$ and, 3.92 $\mu_B$ and -75.49 K for the monoclinic NdPO$_4$, respectively. The negative values of reported T$_\theta$ indicate on strong antiferromagnetic exchange coupling in both phases, which is confirmed by our DFT calculations. In this way, various magnetic configurations of the monoclinic monazite structure were theoretically studied [10]. The investigation of their total energies revealed that the NdPO$_4$ compound is more stable in a type-3 antiferromagnetic ordering (Inset of Fig 1-a) with a direct band gap of 3.4 eV [10].

The evaluation of the magnetocaloric effect particularly in terms of the entropy change (-ΔS) demonstrated the NdPO$_4$ to present a large magnetocaloric effect at very-low temperatures (Fig1-c). In the magnetic field change of 5 T, the anhydrous NdPO$_4$ exhibits a maximum -ΔS of about 19 J/kg K close to 3 K, being significantly higher than its equivalent in the hydrated NdPO$_4$ phase which is about 15 J/kg K in a similar magnetic field. This shown MCE is about 5 times larger than that reported in the NdMnO$_3$ compound which contains the same Nd rare earth element [2, 9]. More importantly, the NdPO$_4$ phases unveil similar or even larger levels of MCE when compared to some reference magnetocaloric materials containing expensive rare earth elements with high magnetic moments such as RMnO$_3$ and RMn$_2$O$_5$ (R = Dy, Ho, Tb) [2].

These findings demonstrate that the NdPO$_4$ compound could be a good candidate for cryogenic magnetic cooling not only because of its good magnetocaloric properties but also on account to the affordability of Neodymium (compared to Dy, Ho, Tb, Er..) and phosphoric acid which is by the way one of the main natural ressource of Morocco which holds more than 70 % of the world's phosphate reserves [10].

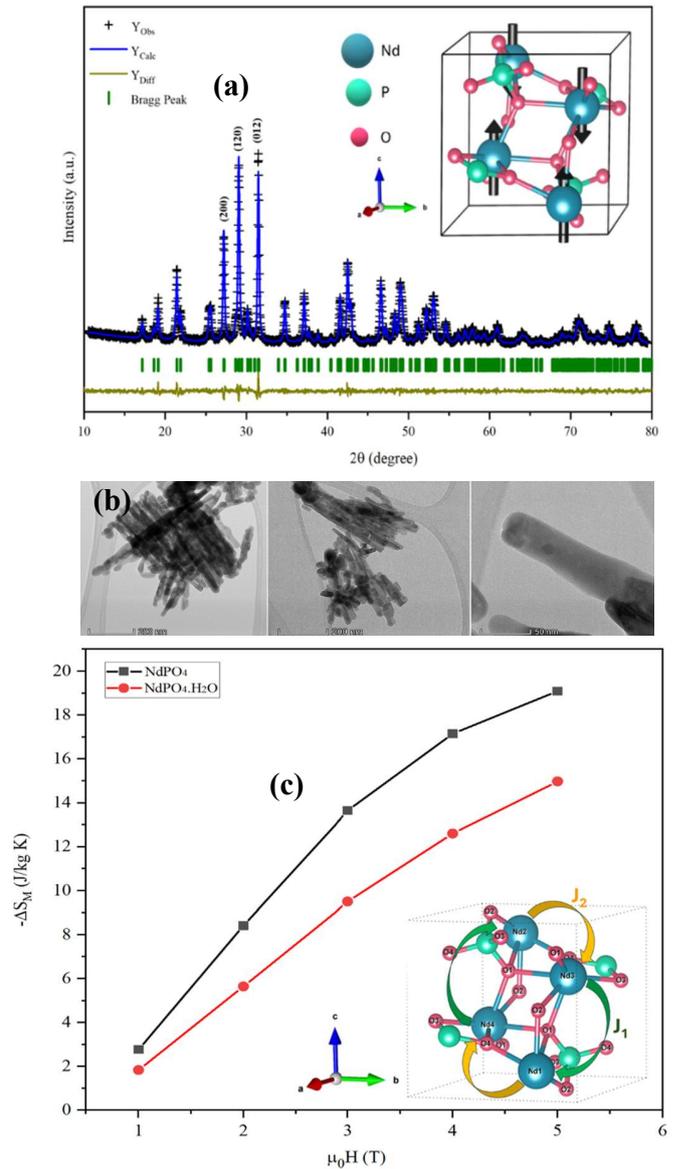

Fig.1. (a) XRD patterns of anhydrous monoclinic NdPO$_4$. (b) TEM images showing nanorods of NdPO$_4$. (c) Magnetic entropy changes as a function of magnetic field at 3 K for anhydrous (square) and hydrated (solid cercle) NdPO$_4$ phases. Inset: exchange interactions in monoclinic NdPO$_4$.

REFERENCES

[1] M. Balli, S. Jandl, P. Fournier, A. Kedous-Lebouc, Appl. Phys. Rev. 4, 021305 (2017).
[2] M. Balli, B. Roberge, P. Fournier and S. Jandl, Crystals 7, 44 (2017).
[3] M. Balli, B. Roberge, S. Jandl, P. Fournier, T. T. M. Palstra, A. A. Nugroho, J. Appl. Phys. 118, 073903 (2015).
[4] M. Balli, S. Mansouri, D. Z. Dimitrov, P. Fournier, S. Jandl, J. Y. Juang, Phys. Rev. Mater. 4, 114411 (2020).
[5] Mohamed Balli, Sohail Ait Jmal, Oumayma Chdil, Sabeur Mansouri, Patrick Fournier, Serge Jandl, Jean-Paul Salvestrini, ACS Appl. Energy Mater. 8, 12415 (2025).

[6] G. A. Gehring and K. A. Gehring, Rep. Prog. Phys. 38, 1 (1975).
[7] G. J. Bowden, Aust. J. Phys. 51, 201 (1998).
[8] E. Palacios et al., Phys. Rev. B 90, 214423 (2014).
[9] L. Attou, K. El Maalam, S-E. Bouzarmine, S. Ait Jmal, Z. Kacemi, M. Ben Ali, S. Naamane, O. Mounkachi, P. Fournier, M. Balli., Materials Today Physics 60, 101989 (2026).
[10] S-E. Bouzarmine, S. Ait Jmal, L. Attou, Z. El Kacemi, S. Mangin, M. Balli, ACS Inor. Chem 65 (5), 3072 (2026).